\documentclass[twocolumn, twoside]{optica-article}

\journal{optica}

\articletype{Research Article}

\begin{document}

\title{Arbitrary electro-optic bandwidth and frequency control in lithium niobate optical resonators}

\author{Jason F. Herrmann\authormark{1,*}, Devin J. Dean\authormark{1}, Christopher J. Sarabalis \authormark{1}, Vahid Ansari \authormark{1}, Kevin Multani \authormark{2}, E. Alex Wollack \authormark{2,4}, Timothy P. McKenna \authormark{3}, Jeremy D. Witmer \authormark{1}, and Amir H. Safavi-Naeini \authormark{1,*}}

\address{\authormark{1}Department of Applied Physics, Stanford University\\
\authormark{2}Department of Physics, Stanford University\\
\authormark{3}Department of Electrical Engineering, Stanford University\\
\authormark{4}Present address: AWS Center for Quantum Computing, San Francisco, CA}

\email{\authormark{*}jfherrm@stanford.edu}
\email{\authormark{*}safavi@stanford.edu}

\begin{abstract}
In situ tunable photonic filters and memories are important for emerging quantum and classical optics technologies. However, most photonic devices have fixed resonances and bandwidths determined at the time of fabrication. Here we present an in situ tunable optical resonator on thin-film lithium niobate. By leveraging the linear electro-optic effect, we demonstrate widely tunable control over resonator frequency and bandwidth on two different devices. We observe up to $\sim50\times$ tuning in the bandwidth over $\sim50$ V with linear frequency control of $\sim230$ MHz/V. We also develop a closed-form model predicting the tuning behavior of the device. This paves the way for rapid phase and amplitude control over light transmitted through our device.
\end{abstract}

\section{Introduction}
Tunable couplers for electromagnetic waves have been proposed and demonstrated for a variety of applications in many regimes, including microwave systems for cavity dumping, catch-and-release, and photon pulse shaping \cite{Pechal2014, Yin2013, Axline2018}, as well as optical systems for spectral compression, pulse storage and shaping, and optical quantum gates \cite{Keller2004, Tian2012, Myilswamy2020, Yanik2004, Yanik2005, Shoman2019, Heuck2019, Heuck2020}. These applications benefit from high-speed amplitude and phase control over the light in the cavity, which are governed respectively by the cavity bandwidth and resonant frequency. Whereas previous demonstrations of tunable resonators on silicon rely on thermo-optic phase shifters or charge injection schemes \cite{Poon2007, Chen2007}, more recent demonstrations on thin-film lithium niobate (TFLN) leverage the linear electro-optic effect (``Pockel's'' effect) to achieve faster phase modulation with low loss \cite{Xue2022,Hwang2022}.

Here, we build on this TFLN platform, demonstrating simultaneous and arbitrary control of both the resonant frequency and bandwidth of optical resonator modes. This dual control is necessary for achieving arbitrary phase and amplitude control over light transmitted through the device. We also derive a closed-form model describing the bandwidth and detuning behavior as a function of the applied voltage. This model is important to calibrate device performance and model its behavior in pulse shaping applications.

\section{Results}
\subsection{Device Geometry and Operation}
\begin{figure}
    \centering
    \includegraphics[width=80mm]{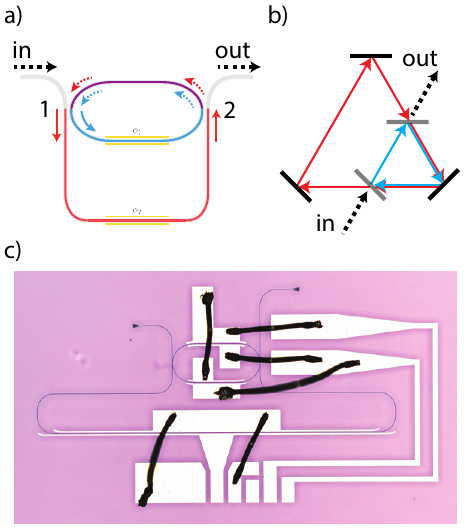}
    \caption{Device geometry and operation. (a) Schematic of the device. Light enters from the left and can couple into the racetrack resonator at points ``1'' and ``2.'' Red and blue coloring represent two independent paths the light can take, with purple corresponding to the light in both paths overlapping. Gold represents electrodes. The light can accumulate a different phase $\phi_1$ or $\phi_2$ in each path. (b) An alternative, free-space optics schematic of the system. Black lines are mirrors, and grey lines are tunable mirrors. Two cavities are formed by the light oscillating in the broad ring and the light in the small cavity. Interference between the light in both cavities varies the coupling in and out of the smaller cavity. (c) Optical microscope image of a similar device, representative of the devices measured in this paper.}
    \label{fig:Fig1}
\end{figure}

Our device comprises a resonator with an output waveguide that incorporates phase-sensitive feedback to tune the coupling rate. It consists of a TFLN ridge-slab waveguide and racetrack resonator atop a sapphire handle. A similar device to those measured in this work is pictured in Fig.\ref{fig:Fig1}c. The ridge waveguide is $\sim1.2$ $\mu$m wide, $\sim300$ nm tall, and rests atop a $\sim200$ nm-thick slab layer. It is designed for single-mode operation at telecommunications wavelengths. We fabricate our device following the techniques developed in \cite{McKenna2020}; waveguides are patterned via an HSQ hardmask, exposed with electron-beam lithography (JEOL JBX-6300FS), and etched using argon ion milling. We pattern Ti:Au electrodes via photolithography and lift-off. Lastly, we make on-chip wirebonds to ensure proper polarity between the electrodes. The electrodes and photonic device are positioned in order to align the applied electric field with the TFLN crystal z-axis, thereby taking advantage of the large $d_{33}$ electro-optic coefficient of LN \cite{Weis1985}.

The racetrack resonator is coupled at two points to the feedback waveguide (Fig.\ref{fig:Fig1}a). Each coupling point acts as a beam splitter. At the first coupling point, the light splits into two paths; one path consists of the bottom half of the racetrack resonator, while the other consists of the feedback waveguide. The electrodes are positioned both across the racetrack resonator and the feedback waveguide, with independent voltage control over the bias applied to each set of electrodes. In each path, the light accumulates a voltage-dependent phase shift,
\begin{equation}\label{eq:phi}
    \phi_i(\omega; V_i) = \beta_{V,i}(\omega; V_i)L_i.
\end{equation}
Here, $i$ refers to either the first (racetrack) or second (waveguide) path, and $\beta_{V,i}(\omega; V)$ is the voltage- and frequency-dependent optical propagation constant in each path for a given mode.  At the second coupling point, the light in both paths recombines with a phase difference $\Delta\phi =\phi_2-\phi_1$. If the light between the two paths is exactly in-phase ($\Delta\phi=0$), the interference is constructive and coupling between the racetrack and the waveguide is enhanced. This appears as a broadening in the mode's linewidth. If the light is perfectly out of phase ($\Delta\phi=\pm\pi)$, the interference is destructive, and the racetrack is completely decoupled from the waveguide. The voltage thus controls the resonator bandwidth. This coupling can be thought of as an Mach-Zehnder Interferometer (MZI) with an output port looped back to an input to form a resonator.

\subsection{Measurement}
\begin{figure}
    \centering
    \includegraphics{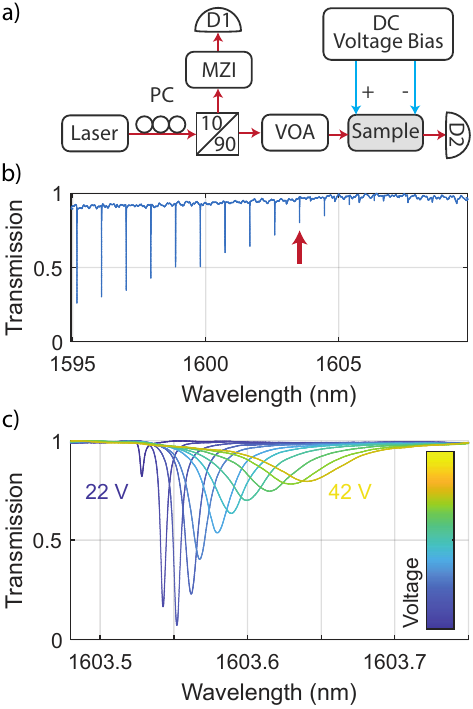}
    \caption{Measurement Apparatus and Optical Transmission. (a) Simplified schematic of our measurement setup. The input laser wavelength is swept across multiple resonances of the device. Two remotely controlled voltage sources are used to simultaneously apply static voltage bias to either the waveguide or racetrack electrodes. (b) An example of a normalized transmission spectrum of the device. In this case, a bias voltage of 22 V is applied to the racetrack electrodes, while a bias of -22 V is applied to the waveguide electrode. The red arrow indicates a particular mode at $\sim 1603.6$ nm, which we track in (c). (c). Overlaid normalized transmission curves for the mode indicated in (b), plotted for every two volts from 22 V to 42 V. The color indicates the magnitude of voltage. The same magnitude but opposite sign is applied to the waveguide electrode versus the racetrack electrodes.}
    \label{fig:Fig2}
\end{figure}

We measure our device using the apparatus schematically depicted in Fig.\ref{fig:Fig2}a. A laser is split into two arms. One passes through the optical device, whereas the other passes through an off-chip MZI for the purpose of calibrating the wavelength axis of the device spectrum (see supplemental section S2). We vary the constant voltage bias across the device electrodes and record the transmission spectrum for each voltage in different bias configurations. Fig.\ref{fig:Fig2}b depicts one such spectrum. In this spectrum, we observe transmission dips corresponding to resonant frequencies of the racetrack. The dip contrast changes with wavelength across the spectrum, corresponding to coupling differences of the modes. In this case, with fixed voltage bias, the change in coupling can be attributed to each mode having a different propagation constant $\beta$, due to changing the mode wavelength, and therefore a different phase difference accumulated in the feedback region of the device. Fig.\ref{fig:Fig2}c presents the evolution of a single mode from this spectrum as the applied voltage is varied for a particular configuration. It can be seen that this mode undergoes a frequency shift, as well as bandwidth tuning from very under-coupled, through critical coupling, to very over-coupled.

\begin{figure}
    \centering
    \includegraphics{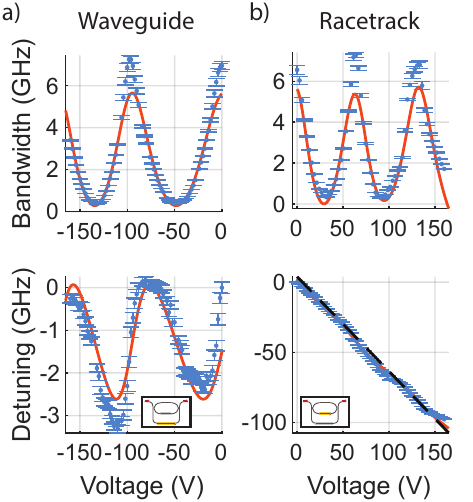}
    \caption{Independent Electrode Bias Calibration. (a) Tuning behavior of the bandwidth (top) and mode detuning (bot) as a function of voltage applied to the waveguide electrode. The inset schematically depicts the device and where the voltage is applied. Blue datapoints represent data with errorbars, and the red line is a fit of our model to the data, weighted by the errorbars. (b) Tuning behavior of the bandwidth (top) and mode detuning (bot) as a function of voltage applied to the racetrack electrode. The inset schematically depicts the device and where the voltage is applied. For this particular device, however, we have an additional racetrack electrode positioned on the upper straight-length of the racetrack. The applied voltage is equivalent across both electrodes. Blue datapoints represent data with errorbars, and the red line is a fit of our model to the data, weighted by the errorbars. Note the large linear detuning component, indicated by the overlaid black dashed line.}
    \label{fig:Fig3}
\end{figure}

In order to calibrate the performance of our device and to demonstrate independent bandwidth and frequency tunability, we repeat these transmission measurements while varying the voltages on the electrodes. First, we apply a voltage only across the feedback waveguide. For each voltage, we fit a Fano resonance to a single mode \cite{Fan2003}. We define ``bandwidth,'' $\kappa(V)$, as the bandwidth parameter of the Fano lineshape, which corresponds to the 3 dB bandwidth of a symmetric Fano (i.e., full-width half-maximum of a Lorentzian). We plot in Fig. \ref{fig:Fig3} the ``detuning,'' $\Delta(\omega)$, between the mode at the current voltage bias and its frequency under zero applied voltage, as well as the bandwidth. Both the bandwidth and detuning exhibit near-sinusoidal oscillations on the order of a few GHz, with a bandwidth extinction ratio (defined as the ratio of maximum bandwidth over minimum bandwidth) of $\sim20$. 

Next, we apply a voltage only to the racetrack electrodes. We again observe near-sinusoidal oscillation in both bandwidth and detuning, but we also see large linear detuning proportional to the applied voltage (Fig. \ref{fig:Fig2}(b,bot)). From this linear fit, we infer the DC electro-optic tuning rate, $g_{EO}/2\pi\approx 640$ MHz/V. We also note that the oscillation in bandwidth in this case arises predominantly from the shifting resonant wavelength in the device. That is, as the resonance frequency of a mode is shifted, its propagation constant varies, thereby altering the phase-difference the mode accumulates between the racetrack and feedback waveguide. 

Table \ref{DeviceParameters} presents the bandwidth extinction ratios and EO tuning strength observed for two different devices (the device measured in Fig. \ref{fig:Fig2}, as well as that presented in supplemental section S1). We observe that the maximum achievable bandwidth scales with the strength of the single-pass power coupling at each racetrack-waveguide coupling point (denoted as ``$r$''). This coupling strength can be increased by making the fabricated racetrack-to-waveguide gap smaller. 
\begin{table}
\begin{center}
\begin{tabular}{c|c|c|c|c}
    \hline
    Device Number & Feedback Length (mm) & $r$ (\%) & $\kappa_{\text{max}}/\kappa_{\text{min}}$ & |$g_{\text{EO,DC}}|/2\pi$ (MHz/V) \\
    \hline
    1 & 2 & 6.0 & 20.2 & 337 \\
    2 & 1.5 & 14.6 & 52.2 & 232\\
\end{tabular} 
\end{center}
\caption{Feedback length is the length of the interferometric waveguide between coupling points, as specified in our device CAD. The term $r$ corresponds to the power coupling across each coupling point, as fit by our model. Importantly, in our model, this term also includes propagation loss factors, and as reported here, assumes no propagation loss. It is therefore a lower-bound on the coupling strength as fit by our model (see supplement section S4). The bandwidth extinction ratio is given by $\kappa_{\text{max}}/\kappa_{\text{min}}$ as observed in experiment. For device 2, we could not measure the maximum bandwidth, and the fit yields an even greater inferred extinction ratio than that reported here. The linear $g_{\text{EO,DC}}$ is inferred from fitting a linear background to the detuning fit of the ring bias calibration, and here is reported for a single electrode on the racetrack. That is, the total detuning of the mode in device 1 is $640$ MHz/V, because it has two electrodes on the racetrack (whereas device 2 has only one electrode; see supplement section S1).}
\label{DeviceParameters}
\end{table}

\subsection{Fitting to a Model}
In order to fit the tuning curves in Figs. \ref{fig:Fig2}, we use scattering matrix theory to derive a full model of the device transmission (see supplemental section S3). This transmission is given by:
\begin{equation}\label{eq:S21}
    S_{21} = \left|\frac{t_1t_2e^{i\phi_{w}(\omega; V_w)}-r_1r_2e^{i\phi_r(\omega; V_r)}-(r_1^2+t_1^2)(r_2^2+t_2^2)e^{i(2\phi_r(\omega; V_r)+\phi_w(\omega; V_w))}}{1-t_1t_2e^{i2\phi_r(\omega; V_r)}+r_1r_2e^{i(\phi_r(\omega; V_r)+\phi_w(\omega; V_w))}}\right|^2
\end{equation}

In equation \ref{eq:S21}, $r_i$ ($t_i$) is the amplitude cross-coupling (transmission) coefficient at each coupling point, $i$. In the case of lossless coupling, $(r_i^2+t_i^2)=1$. The phase $\phi_r(\omega; V_r)$ corresponds to the phase accumulated in a single half of the racetrack resonator (i.e., the phase accumulated between the coupling points in the racetrack), and $\phi_{w}(\omega; V_w)$ is the phase accumulated in the feedback waveguide between the coupling points. Each accumulated phase is in general complex and is given by equation \ref{eq:phi}, where $\beta(\omega; V)$, the complex propagation constant, accounts for both the resonant frequency as well as propagation loss.

The poles of this transfer function correspond to the complex resonances of the system and therefore contain information about both the frequency and bandwidth of each resonance. We solve for the poles by Taylor expanding the denominator close to each resonant frequency, around a frequency $\omega_0$, and set it equal to zero. We make the assumption that higher-order phase derivatives are negligible ($\delta^n\phi_i/\delta\omega^n \ll \left(\delta\phi_i/\delta\omega\right)^n$). This loosely corresponds to the assumption that the magnitude of the group velocity dispersion is much less than the magnitude of the inverse group velocity squared ($(\delta^2\beta/\delta\omega^2) \ll \left(\delta\beta/\delta\omega\right)^2L$). We also assume that the voltage-dependent portion of the phase is not frequency-dependent close to $\omega_0$. In this regime, we arrive at the following form for the expansion (see supplemental section S3). We omit the argument labels of the phases ($\phi_i(\omega_0;V_i)\to\phi_i$) for clarity:
\begin{equation}\label{eq:taylorExpandPoles}
    f(\omega) \approx 1 + \sum_{n=0}^\infty \frac{1}{n!}\left[A\left(\frac{i(L_w+L_r)}{v_g}\right)^ne^{i(\phi_{w}+\phi_{r})}-B\left(\frac{i2L_r}{v_g}\right)^ne^{i2\phi_r}\right]\zeta^n(\omega)\equiv 0
\end{equation}
In the above, we have factored propagation losses, $\gamma_i$ (the complex component of each phase $\phi_i = \phi_i(\omega_0)$) and the amplitude coupling coefficients, $t_i$, $r_i$ into the coefficients $A=r_1r_2\gamma_{w}\gamma_{r}$ and $B = t_1t_2\gamma_{r}^2$. We also define the length of half the racetrack to be $L_r$ and the length of the interferometric waveguide between coupling points to be $L_w$. $v_g$ is the group velocity of the mode, which is taken here to be equivalent in each part of the device. 

We then solve this expression to first order to find the complex pole frequency $\omega_p$, such that the complex detuning, $\zeta(\omega_p) = \omega_p-\omega_0$, from a frequency close to the zero-voltage resonance $\omega_0$, results in $f(\omega_p) = 0$:
\begin{equation}
    \zeta(\omega_p)\equiv \omega_p-\omega_0=-\frac{1+Ae^{i(\phi_{w}+\phi_{r})}-Be^{i2\phi_r}}{A\left(\frac{i(L_w+L_r)}{v_g}\right)e^{i(\phi_{w}+\phi_{r})}-B\left(\frac{i2L_r}{v_g}\right)e^{i2\phi_r}}
    \label{eq:zeta}
\end{equation}
In this way, we obtain the complex pole $\omega_p$ with real and imaginary parts representing the detuning from $\omega_0$ and half of the linewidth $\kappa$. The full model, its fit parameters, and additional details on the fitting procedure are presented in supplemental section S4. The minimum of the bandwidth fit corresponds to the completely decoupled resonant linewidth ($\kappa_i$), whereas the maximum corresponds to the maximally coupled linewidth. 

\subsection{Arbitrary Bandwidth and Frequency Control}
\begin{figure}
    \centering
    \includegraphics{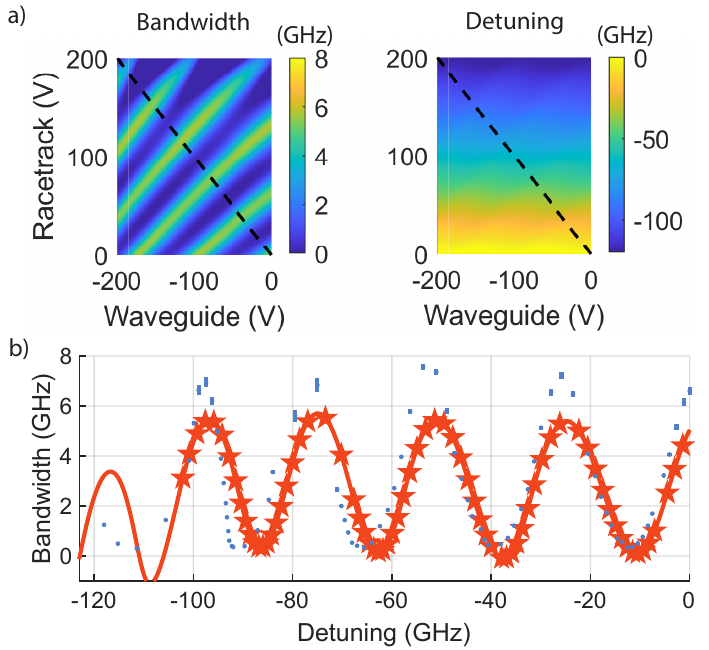}
    \caption{Arbitrary bandwidth and frequency control. (a) Simulations of the expected bandwidth (left, $\kappa$) and detuning (right, $\Delta$) of the $\lambda \sim 1603.6$ nm resonance as a function of applied voltage across the racetrack (y-axis) and waveguide (x-axis). The simulation consists of the predictions from our closed form model, using the results of the fits in Fig. \ref{fig:Fig3}. The dashed black line indicates the voltages over which we measure actual data in (b). (b) Comparison of measured data (blue datapoints) and prediction from our model (red line) for a ``dual bias'' configuration. The stars indicate predictions for the same voltage combinations as those used for the measurement (blue datapoints). We plot the fit bandwidth against detuning of the mode for each applied voltage. In this case, equal magnitude voltage is applied to both the racetrack electrodes and the waveguide electrode, but with opposite polarity. Error bars indicate a 10\% change in the fit error of a Fano fit applied to the resonance from which we extract bandwidth and resonant frequency. The detuning is measured relative to the zero-voltage resonant frequency. In many cases, the error bars are so small they are blocked by the plotted datapoint.} 
    \label{fig:Fig4}
\end{figure}
In order to achieve arbitrary control over bandwidth and frequency of the mode, we can apply bias simultaneously to both the waveguide and racetrack electrodes. We predict the behavior of the device by feeding the fit parameters from Figs. \ref{fig:Fig2} into our model, eqs. \ref{eq:zeta}, S15-17, and simulating the expected bandwidth and detuning of the mode as functions of the applied racetrack and waveguide voltages. The full list of parameters for the model are presented in supplemental section S4. The simulated results are presented in Fig. \ref{fig:Fig4}. As expected, we observe a periodic modulation of both the bandwidth and detuning with voltage. For fixed waveguide voltage but varied racetrack voltage (i.e., a vertical slice in Fig. \ref{fig:Fig4}(a)), we observe a strong linear detuning in the resonant frequency and a modulation in bandwidth. For a fixed racetrack voltage but varied waveguide voltage (i.e., a horizontal slice), we observe complete modulation between maximum and minimum coupling, as well as a modulation in detuning. Therefore, by varying the applied voltages, we can access any bandwidth and frequency within this simulated 2D space. Furthermore, we can compensate for any undesired detuning modulation arising from the waveguide bias with linear frequency shift controlled by the racetrack bias.

We verify this simulated result experimentally by applying voltages corresponding to the dashed diagonal lines in Fig. \ref{fig:Fig4}(a). We apply voltages of equal magnitude but opposite polarity to the waveguide and racetrack electrodes. Sweeping this voltage from 22 V to 42 V, we take repeated transmission measurements and plot the bandwidth versus detuning of the $1603.6$ nm mode for each voltage (Fig. \ref{fig:Fig4}(b)). We observe a close match between our measured device behavior and the simulated predictions from Fig. \ref{fig:Fig4}(a).

\section{Discussion}
\subsection{Experimental Considerations}
Overall, we have demonstrated independently tunable control over both the resonant frequencies and bandwidths of modes of an integrated optical resonator. The achievable frequencies are limited only by how much bias voltage can be applied to the racetrack. The required voltage for a desired phase shift can be reduced by making the straight lengths of the racetrack (and therefore the electrodes) much longer, and also by making the FSR of the racetrack shorter (so that one could operate on a different mode to achieve a frequency change). The maximum bandwidth achievable in this device is limited by the amount of power exchange between the racetrack and the waveguide at each coupling point. We could increase coupling in the future by either reducing the waveguide-to-racetrack coupling gap or moving to longer coupling regions, as in directional or pulley couplers.

Furthermore, when fabricating the device, the physical symmetry of the two coupling points is important for ensuring the maximum bandwidth extinction ratio. If one coupler allows for more power transfer than the other, this will appear as a loss channel on the racetrack resonator, increasing the minimum resolvable bandwidth in the system (see supplemental section S6). This is similar to the behavior in an MZI in which asymmetric couplers would cause incomplete extinction of the light in each arm.

In the device exhibited in this work, our wavelength calibration is also tricky. The MZI we used at the time of measurement has an FSR of approximately $7.65$ GHz, which is much larger than our minimum resonance bandwidth. By calibrating to the peaks, valleys, and zero-crossings of the MZI, we are able to obtain wavelength references at every $\sim1.9$ GHz, and we interpolate the wavelengths in between. For our supplemental device 2, we used an MZI with a much narrower FSR ($\sim 325$ MHz) and a much more accurate calibration scheme. Both calibration algorithms work best with a smooth MZI transmisison. Our MZI transmission data features a regular fast ripple, likely coming from some equipment in the laser path. We smooth the MZI transmission data prior to calibration to eliminate the ripple. Furthermore, the device 1 transmission is measured through a preamplifier with an active low-pass filter (this filter is turned off when measuring supplemental device 2). However, we note that this filtering only serves to \textit{increase} our minimum-resolvable bandwidth, thereby placing a \textit{lower-bound} on the bandwidth extinction ratio we measure.

\subsection{Data Considerations}
In Fig. \ref{fig:Fig2}, we observe near-sinusoidal tuning behavior in bandwidth and detuning. Previous demonstrations of feedback-coupled waveguides depicted nearly exact sinusoidal tuning behavior \cite{Chen2007,Poon2007}. We attribute the asymmetry we observe in the oscillation to the large length mismatch between arms of the feedback coupling region. In order to reduce the voltage necessary to achieve a complete phase shift in the waveguide, we choose to make the waveguide very long. In future device iterations, we would make the racetrack straight length similarly long, thereby mitigating the issues of the length mismatch. We investigate this imbalance more thoroughly via simulation in supplemental section S6.

We also notice asymmetry in the voltage required for a complete period of bandwidth tuning in Fig. \ref{fig:Fig2}(a). That is, the $V_{\pi}$ of our device is greater at low voltages. We attribute this difference to charge carriers in TFLN. Along these lines, we expect the tuning period to shift with optical power, which would alter the rate of free carrier generation. We leave further study of these effects to studies focused on modulators.

Lastly, we note that our modeling required a first-order approximation for the complex poles in the device to arrive at semi-analytic expressions for the linewidth and detuning. Therefore, it is only valid in a narrow region around each resonant frequency, and outside of this region, the model may predict non-physical behavior. An example of this is shown in Fig. \ref{fig:Fig4}b. As the applied voltage grows very large, (e.g., beyond $\sim120$ V), the predicted bandwidth of the mode can be negative. We believe this stems from the local nature of our Taylor expansion and the fact that a high voltage leads to large linear detuning of the mode. However, this warrants further theoretical and experimental investigation. 

\subsection{Conclusion}
We have presented an approach to tune the transmission properties of integrated photonic resonators. By leveraging the fast electro-optic effect in TFLN, our approach will also enable optical pulse shaping, and photon catch and release experiments important for emerging quantum networks. By developing a closed-form model to calibrate and predict tuning behavior of this device, we pave the way for these experiments and others that require precise models and repeatable operation.

\section{Backmatter}


\begin{backmatter}
\bmsection{Funding}
Placeholder text. This will be filled automatically via the submission.

\bmsection{Acknowledgments}
 We thank NTT Research for their financial and technical support. We thank the United States government for their support through the Department of Energy Grant No. DE-AC02-76SF00515, the Defense Advanced Research Projects Agency (DARPA) LUMOS program (Grant No. HR0011-20-2-0046), the DARPA Young Faculty Award (YFA, Grant No. D19AP00040),  the U.S. Department of Energy (Grant No. DE-AC02-76SF00515) and Q-NEXT NQI Center, and the U.S. Air Force Office of Scientific Research MURI grant (Grant No. FA9550-17-1-0002). JFH and DJD would like to acknowledge support from the NSF GRFP (No. DGE-1656518). VA acknowledges support from the Stanford Q-FARM Bloch Fellowship Program. KM acknowledges support from the Natural Sciences and Engineering Research Council of Canada (NSERC). Part of this work was performed at the Stanford Nano Shared Facilities (SNSF), supported by the National Science Foundation under award ECCS-2026822. Work was performed in part in the nano@Stanford labs, which are supported by the National Science Foundation as part of the National Nanotechnology Coordinated Infrastructure under award ECCS-2026822. 

\bmsection{Disclosures}
The authors declare no conflicts of interest.

\bmsection{Data Availability Statement}
All data generated in this study are available from the authors upon reasonable request.

\bmsection{Supplemental document}
A supplemental document containing additional device analysis and discussion has been submitted with this manuscript.

\end{backmatter}

\clearpage

\title{Arbitrary electro-optic bandwidth and frequency control in lithium niobate optical resonators: Supplementary Information}

\section{Additional Device Data: Device 2}\label{sec:supp_d2_data}
\begin{figure}[h]
    \centering
    \includegraphics[]{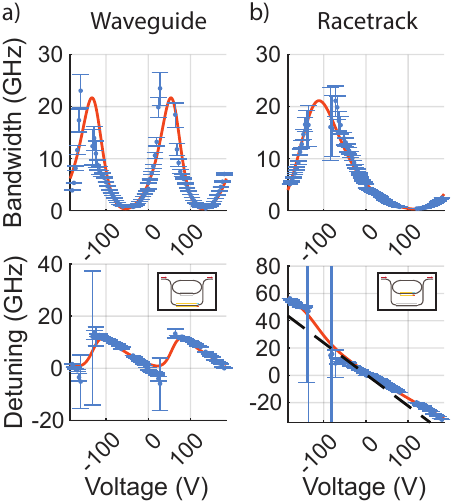}
    \caption{Tuning curves for device 2 under applied bias configurations. (a) Bias applied along waveguide electrode (``electrode 1''). Bandwidth (top) and detuning (bot) versus applied voltage. Blue is datapoints, red is fit. (b) Bias applied along racetrack electrode (``electrode 2''). Bandwidth (top) and detuning (bot) versus applied voltage. Blue is datapoints, red is fit. The black dashed line indicates the linear tuning component of the resonant frequency from the applied racetrack voltage. Note that unlike device 1, device 2, shown here, only has a single electrode on the racetrack, leading to approximately half the linear tuning range for the same applied racetrack voltage as compared to device 1. The error bars are determined using the technique discussed in section~\ref{supp:sec_error}.}
    \label{fig:FigS1_a}
\end{figure}

Here we present the data and tuning curves for device 2. This device is fabricated with a shorter interferometric waveguide and smaller racetrack-to-waveguide coupling gaps relative to device 1. Therefore, we expect larger bandwidth tunability on device 2, which we observe. The tuning curves are depicted in Fig.~\ref{fig:FigS1_a}. We fit our model (derived in section \ref{sec:supplement_modeling}) using the techniques in section \ref{sec:supplement_fitting}. This same fitting technique is applied to the device 1 data in the main text.

\begin{figure}[h]
    \centering
    \includegraphics[]{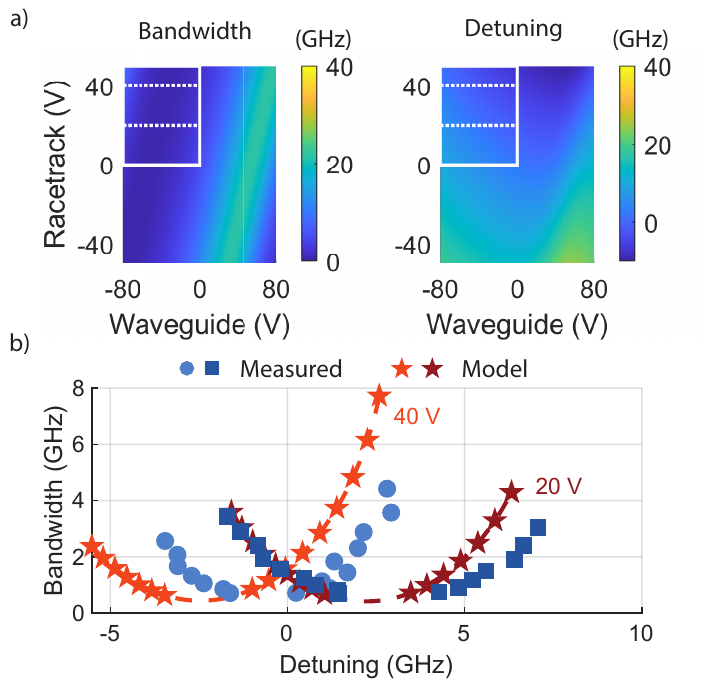}
    \caption{Dual bias modeling and prediction for device 2. (a) Simulated bandwidth and detuning for applied racetrack and waveguide voltage combinations. The white-outlined box indicates a particular region of interest, and the dashed white lines indicate the voltage combinations used for the data in (b). (b) Measured bandwidth versus detuning of modes for the indicated fixed racetrack voltages of $20$ V and $40$ V, highlighted by the dashed white lines in (a). The waveguide voltage is swept from $0$ V to $-80$ V. The ``model'' lines indicate predictions of the expected tuning behavior, and stars indicate the predictions at the same voltage combinations used for the measured data.}
    \label{fig:FigS1_b}
\end{figure}

In Fig.~\ref{fig:FigS1_b}, we plot the predicted dual-bias tuning behavior, using the parameters returned from the fits in Fig.~\ref{fig:FigS1_a}. We measure and plot the bandwidth versus detuning for modes under fixed racetrack voltages while sweeping the applied waveguide voltage. Our model reasonably predicts the expected tuning behavior for these voltage combinations, as shown in Fig.~\ref{fig:FigS1_b}(b).


\section{Wavelength Calibration}
We calibrate the wavelength axis of the data in two different ways, depending on the data. For device 1, we use an MZI with a larger FSR compared to that used for device 2. At the time of data collection for device 1 we also used a low-pass filter on our preamplifier. The effect of this filter, however, would be to increase the minimum resolvable bandwidth on the mode, so our reported values for minimum bandwidth and bandwidth extinction ratio are conservative values. Given the similarity between this value and the minimum bandwidth for device 2, for which we used DC coupling to the preamplifier (bypassing any filtering), we suspect the filtering has little effect on the measured results. The MZI used for measuring device 1 has an FSR of approximately $7.65$ GHz. Using all peaks, valleys, and zero-crossings as calibration points, this yields a wavelength reference of roughly every $1.91$ GHz. This is much larger than the minimum bandwidth in our system. On the other hand, for device 2, we use an MZI with an FSR of $\sim325$ MHz, which is comparable to the minimum bandwidth measured in the system. For device 2, we also utilize all phase information from the MZI, ensuring more accurate overall wavelength calibration for these datasets.

\section{Scattering Matrix Model}\label{sec:supplement_modeling}
\subsection{Deriving the scattering matrix}
We can represent our device as four connected regions. Each region is described by a two port scattering matrix. We refer to the top path (the racetrack) as path $1$ and the bottom path (the feedback waveguide) as path $2$. The input and output to each segment are given as $\vec{a}_{i,j}$ and $\vec{b}_{i,j}$ respectively, where $i$ indexes the path, and $j$ indexes the segment. We define the following four scattering matrices:
\begin{align}
    \begin{pmatrix}
        b_{1,1} \\
        b_{2,1}
    \end{pmatrix} &= 
    \begin{pmatrix}
        t_1 & ir_1 \\
        ir_1 & t_1
    \end{pmatrix}
    \begin{pmatrix}
        a_{1,1}\\
        a_{2,1}
    \end{pmatrix}\\ 
    \begin{pmatrix}
        b_{1,2} \\
        b_{2,2}
    \end{pmatrix} &= 
    \begin{pmatrix}
        e^{i\phi_r} & 0 \\
        0 & e^{i\phi_{w}}
    \end{pmatrix}
    \begin{pmatrix}
        a_{1,2}\\
        a_{2,2}
    \end{pmatrix}\\
    \begin{pmatrix}
        b_{1,3} \\
        b_{2,3}
    \end{pmatrix} &=
    \begin{pmatrix}
        t_2 & ir_2 \\
        ir_2 & t_2
    \end{pmatrix}
    \begin{pmatrix}
        a_{1,3}\\
        a_{2,3}
    \end{pmatrix}\\    
    \begin{pmatrix}
        b_{1,4} \\
        b_{2,4}
    \end{pmatrix} &=
    \begin{pmatrix}
        e^{i\phi_r} & 0\\
        0 & 1
    \end{pmatrix}
    \begin{pmatrix}
        a_{1,4}\\
        a_{2,4}
    \end{pmatrix}
\end{align}

In the above, we have the following definitions:
\begin{itemize}
    \item $t_i$ ($r_i$): transmission (cross-coupling) at coupling point $i$.
    \item$\phi_r\equiv\phi_r(\omega; V_r)$: phase accumulated in one half of the racetrack.
    \item $\phi_{w}\equiv\phi_w(\omega; V_w)$: phase accumulated between coupling points in the interferometric waveguide.
\end{itemize}

We note that the input to each section is equivalent to the output of the previous section; this corresponds to the condition that $a_{i,j+1}=b_{i,j}$. We are interested in the output $b_{1,4}$, so we write:
\begin{equation}
    \begin{pmatrix}
        b_{1,4} \\
        b_{2,4}
    \end{pmatrix} =
    \begin{pmatrix}
        e^{i\phi_r} & 0\\
        0 & 1
    \end{pmatrix}
    \begin{pmatrix}
        t_2 & ir_2 \\
        ir_2 & t_2
    \end{pmatrix}
     \begin{pmatrix}
        e^{i\phi_r} & 0 \\
        0 & e^{i\phi_{w}}
    \end{pmatrix}
     \begin{pmatrix}
        t_1 & ir_1 \\
        ir_1 & t_1
    \end{pmatrix}
    \begin{pmatrix}
        a_{1,1}\\
        a_{2,1}
    \end{pmatrix}
\end{equation}

\begin{figure}
    \centering
    \includegraphics{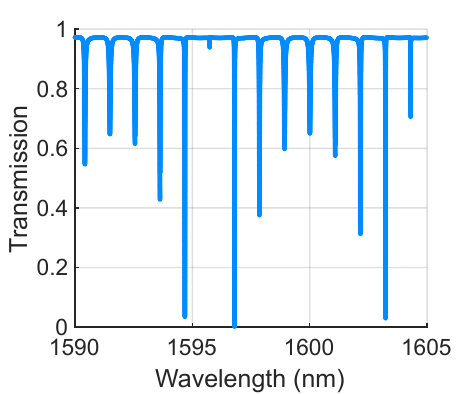}
    \caption{Simulated transmission through device under no voltage bias. Here we plot a simulation of the device, using the resulting fit parameters from the tuning curves of device 1 in the main text. We assume a propagation loss of $5$ dB/cm and plot the $S_{21}$ (eq. \ref{eq:S21}) versus wavelength. This simulated spectrum is qualitatively similar to that of the fabricated device.}
    \label{fig:supplement_s2_spectrum}
\end{figure}

We assert the resonance condition, $a_{1,1} = b_{1,4}$, and then solve for $b_{2,4}$ under a normalized input, $a_{2,1} = 1$, thereby obtaining $S_{21}$, the transmitted light for waveguide input:
\begin{equation}\label{eq:S21}
    S_{21} = \left|\frac{t_1t_2e^{i\phi_{w}(\omega; V_w)}-r_1r_2e^{i\phi_r(\omega; V_r)}-(r_1^2+t_1^2)(r_2^2+t_2^2)e^{i(2\phi_r(\omega; V_r)+\phi_w(\omega; V_w))}}{1-t_1t_2e^{i2\phi_r(\omega; V_r)}+r_1r_2e^{i(\phi_r(\omega; V_r)+\phi_w(\omega; V_w))}}\right|^2
\end{equation}
We simulate eq. \ref{eq:S21} to confirm its validity and plot the results in Fig.\ref{fig:supplement_s2_spectrum}.

\subsection{Complex pole analysis}\label{supp:denomExpansion}
As discussed in the main text, we can obtain information about the resonant frequencies and bandwidths in the system from the complex poles of the transfer function, eq. \ref{eq:S21}. We label the denominator as $f(\omega)$ and define $A=r_1r_2$, $B=t_1t_2$. We also omit the arguments $\omega$ and $V$ from the $\phi_i$ terms for clarity:
\begin{equation}
    f(\omega) = 1-Be^{i2\phi_r}+Ae^{i(\phi_r+\phi_w)}
\end{equation}
Performing a first-order Taylor expansion around the resonant frequency, $\omega_0$:
\begin{equation}
    f(\omega) = f(\omega_0) + f'(\omega_0)(\omega-\omega_0) + \frac{1}{2}f''(\omega_0)(\omega-\omega_0)^2 + ...
\end{equation}
Taking the first few derivatives:
\begin{equation}
\begin{split}
    f'(\omega_0) =& -i2Be^{i2\phi_r}\frac{d\phi_r}{d\omega}+iAe^{i(\phi_r+\phi_w)}\left(\frac{d\phi_r}{d\omega} + \frac{d\phi_w}{d\omega}
    \right)\\
    f''(\omega_0) = & -(i2)B\left[i2e^{i2\phi_r}\left(\frac{d\phi_r}{d\omega}\right)^2 + e^{i2\phi_r}\left(\frac{d^2\phi_r}{d\omega^2}\right)\right]\\
    & +iA\left[ie^{i(\phi_r+\phi_w)}\left(\frac{d\phi_r}{d\omega}+\frac{d\phi_w}{d\omega}\right)^2 + e^{i(\phi_r+\phi_w)}\left(\frac{d^2\phi_r}{d\omega^2}+\frac{d^2\phi_w}{d\omega^2}\right)\right]
\end{split}
\end{equation}
From the definition for $\phi_i$, we have:
\begin{equation}\label{eq:defPhi_expand}
    \phi_i = \beta_{0,i}(\omega)L_i + \Delta\beta_{0,i}(V_i)L_i
\end{equation}
In eq. \ref{eq:defPhi_expand}, $\beta_{0,i}$ is the propagation constant in segment $i$ under zero applied voltage, and $\Delta\beta_{0,i}(V_i)$ is the shift in the propagation constant resulting from applied voltage, $V_i$. We obtain the derivative (dropping the arguments for clarity):
\begin{equation}
    \frac{\delta\phi_i}{\delta\omega} = \frac{\delta\beta_{0,i}}{\delta\omega}L + \frac{\delta\Delta\beta(V)}{\delta\omega}L
\end{equation}

We apply the following three assumptions:
\begin{enumerate}
    \item Higher-order derivatives of the phase, $\phi_i$, are negligible:
    \[\frac{\delta^2\phi_i}{\delta\omega^2} \ll \left(\frac{\delta\phi_i}{\delta\omega}\right)^2\]
    \item The voltage-dependent change in propagation constant is independent of frequency close to the resonance, $\omega_0$:
    \[\frac{\delta\Delta\beta_{0,i}(V_i)}{\delta\omega}\approx 0\]
    \item The group velocity, $v_g$, is approximately equal between the racetrack and waveguide paths:\[\frac{\delta\omega}{\delta\beta_{0,r}} \approx \frac{\delta\omega}{\delta\beta_{0,w}}\equiv v_g\]
\end{enumerate}

From assumption 2, we define the $n$'th derivative of $f(\omega)$ as follows:
\begin{equation}
f^n(\omega_0)\approx-(i2)^nBe^{i2\phi_r}\left(\frac{\delta\phi_r}{\delta\omega}\right)^n+(i)^nAe^{i(\phi_r+\phi_w)}\left(\frac{\delta\phi_r}{\delta\omega}+\frac{\delta\phi_w}{\delta\omega}\right)^n
\end{equation}

From assumption 1, we can rewrite this:
\begin{equation}
f^n(\omega_0)\approx-(i2)^nBe^{i2\phi_r}\left(\frac{\delta\beta_{0,r}}{\delta\omega}\right)^nL_r^n+(i)^nAe^{i(\phi_r+\phi_w)}\left(\frac{\delta\beta_{0,r}}{\delta\omega}L_r+\frac{\delta\beta_{0,w}}{\delta\omega}L_w\right)^n
\end{equation}

Lastly, from applying assumption 3, we can write the Taylor expansion as:
\begin{equation}
f(\omega)\approx 1+\sum_{n=0}^{\infty}\frac{1}{n!}\left[A\left(\frac{i(L_r+L_w)}{v_g}\right)^ne^{i(\phi_r+\phi_w)}-B\left(\frac{i2L_r}{v_g}\right)^ne^{i2\phi_r}\right]\zeta^n(\omega)
\end{equation}

Here $\zeta(\omega)=\omega-\omega_0$ and is in general complex.

We can now solve for the pole, $\omega_p$, which corresponds to setting $f(\omega_p)=0$ and solving for $\zeta(\omega_p)$:

\begin{equation}\label{eq:zeta}
    \zeta(\omega_p)=-\frac{1+Ae^{i(\phi_r+\phi_w)}-Be^{i2\phi_r}}{A\left(\frac{i(L_r+L_w)}{v_g}\right)e^{i(\phi_r+\phi_w)}-B\left(\frac{i2L_r}{v_g}\right)e^{i2\phi_r}}
\end{equation}

The real and imaginary parts of this expression yield the half-bandwidth and detuning of the pole from the frequency $\omega_0$:
\begin{align}
    \label{eq:fit_k_condition}
    \kappa(V)&=-2\mathbf{Im}[\zeta(\omega_p)]\\
    \label{eq:fit_D_condition}
    \Delta(V)&=\mathbf{Re}[\zeta(\omega_p)]
\end{align}

\section{Fitting the Model to the Data}\label{sec:supplement_fitting}
We fit the data using the model given by eqs. \ref{eq:zeta}-\ref{eq:fit_D_condition} in the following way:
\begin{enumerate}
    \item Fit the model to the waveguide-bias-only data (main text Fig.~3a, supplemental Fig.~\ref{fig:FigS1_a}a.
    \item Feed the results from (1) into fitting the racetrack-bias-only data (main text Fig.~3b, supplemental Fig.~\ref{fig:FigS1_a}b).
    \item Use the results from (1) and (2) to predict the results under dual arbitrary bias, and compare to measured data (main text Fig.~4, supplemental Fig.~\ref{fig:FigS1_b}).
\end{enumerate}

We adapt eqs.~\ref{eq:zeta}-\ref{eq:fit_D_condition} into a 12-parameter model, given as follows with fit parameters $p_j$:
\begin{align}
    \label{eq:zeta_fit_model}
    \zeta(\omega_p;V_r;V_w)&=-\frac{1+p_2e^{i(\phi_r+\phi_w)}-p_3e^{i2\phi_r}}{p_2\left(\frac{i(L_r+L_w)}{p_1}\right)e^{i(\phi_r+\phi_w)}-p_3\left(\frac{i2L_r}{p_1}\right)e^{i2\phi_r}}\\
    \label{eq:kappa_fit_model}
    \kappa(\omega_p;Vr;V_w)&=2\textbf{Im}[\zeta(\omega_p;V_r;V_w)]\\
    \label{eq:delta_fit_model}
    \Delta(\omega_p;Vr;V_w)&=-\textbf{Re}[\zeta(\omega_p;V_r;V_w)]+p_{11}V_r+p_{12}
\end{align}
In general, the phase $\phi_i = \beta_{0,i}(\omega;V_i)L_i$ is complex, and the imaginary part of the phase yields propagation loss. We lump this loss term into the prefactors, $p_2$ and $p_3$.
Eq.~\ref{eq:zeta_fit_model} also incorporates the following definitions:
\begin{itemize}
    \item $\phi_w = \beta_{V,w}(\omega;V_w)L_w = p_4+p_6V_r+p_7V_w+p_8V_w^2$
    \item $\phi_r = \beta_{V,r}(\omega;V_r)L_r = p_5+p_9V_r+p_{10}V_r^2$
    \item $V_r$ = phase applied to one half of the racetrack
    \item $V_w$ = phase applied to the interferometric waveguide    
\end{itemize}
The fit parameters, $p_j$, are defined as:
\begin{itemize}
    \item $p_1=v_g$ = Group velocity, fixed as a parameter and not allowed to vary in the fit. This group velocity is computed from the $0$ V spectrum of each device around the mode of interest, assuming the length of the racetrack to be that specified in CAD. We disregard any doublet modes in the spectrum when computing this $v_g$.
    \item $p_2 = A = r_1r_2\gamma_r\gamma_w$ = power cross-coupling at each coupling point, multiplied by the propagation losses in each part of the device, $\gamma_i$. If the coupling points are symmetric, $r_1=r_2$. $A$ is a lower-bound on the power coupling across each point.
    \item $p_3 = B = t_1t_2\gamma_r^2$ = power transmission at each coupling point, multiplied by the total propagation loss in the racetrack. If the coupling points are symmetric, $t_1 = t_2$. $B$ is a lower-bound on the transmitted power past each coupling point.
    \item $p_4=\beta_{0,w}(\omega_0)$ = the zero-voltage propagation constant of the original resonance frequency, $\omega_0$, in the interferometric waveguide.
    \item $p_5=\beta_{0,r}(\omega_0$) = the zero-voltage propagation constant of the original resonance frequency, $\omega_0$, in the raceetrack.
    \item $p_6$ = the change in the waveguide propagation phase, $\Delta\beta_{0,w}L_w$ that comes from shifting the resonance frequency by applying voltage to the racetrack.
    \item $p_7$ = voltage-dependent shift in the waveguide propagation phase, $\Delta\beta_{0,w}L_w$, arising from voltage applied across the waveguide
    \item $p_8$ = quadratic shift in the waveguide propagation phase, which we attribute to carrier screening effects
    \item $p_9$ = voltage-dependent shift in the racetrack propagation phase, $\Delta\beta_{0,r}L_r$ arising from voltage applied across the racetrack
    \item $p_{10}$ = quadratic shift in the racetrack propagation phase, which we attribute to carrier screening effects
    \item $p_{11}$ = linear shift in the resonant frequency $\omega_0$ arising from voltage applied to the racetrack
    \item $p_{12}$ = linear offset in the resonant frequency $\omega_0$
\end{itemize}

We identify the optimal fit with a particle swarm optimization (PSO), based on the cost function, $F$, given in eq.~\ref{eq:S_costFunction}. $F$ combines the sum of least-squares errors for both the bandwidth and detuning data over all voltages, weighted by the average of the upper and lower error bounds at each data point, $\epsilon$. We determine the error bounds using the method in section \ref{supp:sec_error}. $F$ is defined as:
\begin{equation}
    \label{eq:S_costFunction}
    F = \sum_i\frac{1}{\epsilon_{i,\text{B}}}|\kappa(\omega,V_w,V_r)-y_{i,\text{B}}|^2 + \sum_i\frac{1}{\epsilon_{i,\text{D}}}|\Delta(\omega,V_w,V_r)-y_{i,\text{D}}|^2 
\end{equation}

In eq. \ref{eq:S_costFunction}, the subscripts $B,D$ refer to the ``(B)andwidth'' and ``(D)etuning'' of each datapoint. Furthermore, device 2 in supplemental section \ref{sec:supp_d2_data} only has a single racetrack electrode (as compared to the two racetrack electrodes on device 1 in the main text). Therefore, when fitting device 2, we modify the $2\phi_r$ phase term to be $\phi_r + p_5$, thereby only including the voltage-dependent phase shift on one half of the racetrack.

\begin{table} 
\centering
\begin{tabular}{c|c|c|c|c}
    \hline
    Parameter & Device 1 & Device 2 \\
    \hline
    $p_1$ & $1.336\times 10^8$ m/s & $1.319\times 10^8$ m/s\\
    $p_2$ & $0.0602$ & $0.146$\\
    $p_3$ & $0.9281$ & $0.786$\\
    $p_4$ & $-3.1416$ & $0.97$\\
    $p_5$ & $-3.1081$ & $3.012$\\
    $p_6$ & $-0.0953$ & $-0.0161/\text{V}$\\
    $p_7$ & $0.0563/\text{V}$ & $0.0340/\text{V}$\\
    $p_8$ & $-9.2060\times 10^{-5}/\text{V}^2$ &  $4.6345\times 10^{-6}/\text{V}^2$\\
    $p_9$ & $0.0015/\text{V}$ & $2.0588\times 10^{-4}/\text{V}$\\
    $p_{10}$ & $-1.4253\times 10^{-5}/\text{V}^2$& $-2.0102\times 10^{-6}/\text{V}^2$\\
    $p_{11}$ & $-674.6$ MHz/V& $-231.6$ MHz/V\\
    $p_{12}$ & $3.9117$ GHz & $0.7085$ GHz\\
    $L_r$ & $613.3$ $\mu$m & $613.3$ $\mu$m \\
    $L_w$ & $2$ mm & $1.5$ mm\\
    \hline
\end{tabular}
\caption{Fit parameters from the particle swarm optimization. $L_r$ is the half-length of the racetrack, and $L_w$ is the length of the interferometric waveguide between coupling points. The linear detuning $p_{11}$ yields the value of $g_{\text{EO,DC}}/2\pi$ reported in the main text.}
\label{tab:fit_parameters}
\end{table}

\subsection{Fitting Waveguide Bias}
When fitting the waveguide-bias-only data, we set $V_r$ to be zero, assuming fluctuations in the racetrack voltage supply are negligible. This reduces the number of fit parameters in our model. We also set the linear detuning rate to be zero ($p_{11} = 0$), but we allow the detuning offset, $p_{12}$, to vary.

\begin{figure}[h]
    \centering
    \includegraphics{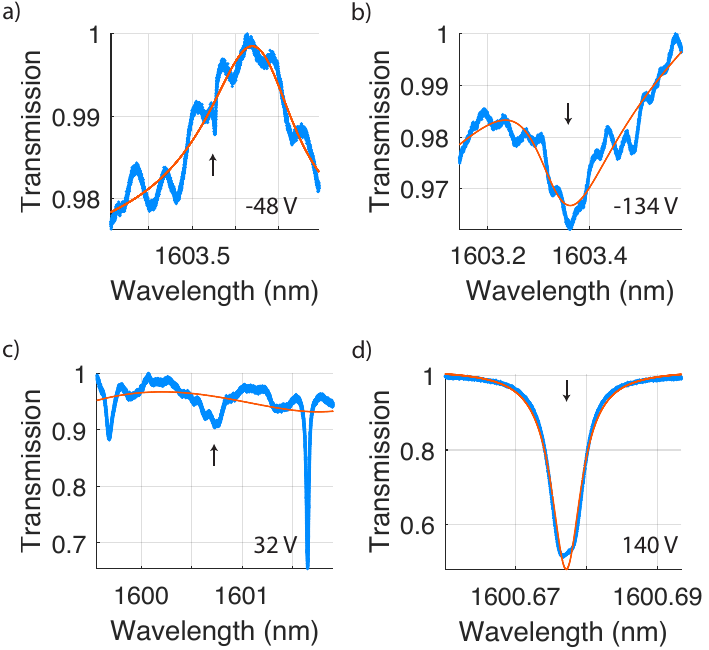}
    \caption{Examples of fits rejected for being ``bad'' fits. Blue is data, red is the fit returned by our algorithm, and the black arrows indicate approximate mode locations. All data corresponds to waveguide-only bias at the indicated voltages. (a) Device 1, under-coupled. The bandwidth of the mode is so small that it blends with the background and is difficult to fit. (b) Device 1, over-coupled. The bandwidth of the mode is so large that it blends with the background. (c) device 2, over-coupled. (d) device 2, doublet. Here we see a doublet beginning to form due to the coupling of the mode of interest with another mode in the background, which is only visible as the coupling of the mode to the waveguide decreases. We discard any fits corresponding to modes that have a clearly visible doublet.}
    \label{fig:supplement_bad_fits}
\end{figure}

As explained in the main text, the dataset consists of the extrapolated bandwidth and detuning of the resonance from Fano fits of the mode spectrum at each applied bias. Before fitting the tuning curves, we filter this data, removing any datapoints for which the Fano fit is clearly inaccurate or appears untrustworthy. This is most relevant at datapoints with very high or very small bandwidth, where the mode blends with the spectrum background and is no longer a true Fano resonance (Fig~\ref{fig:supplement_bad_fits}). We also disregard any datapoints that exhibit an error in the wavelength calibration.

\subsection{Fitting Racetrack Bias}
We repeat for the racetrack bias the same process as for fitting waveguide bias data, feeding forward and fixing parameters that are mutual to both datasets (i.e., $p_1$-$p_5$). We also set all $V_w$ coefficients to zero to reduce the number of fit parameters, and we allow $p_{11}$ and $p_{12}$ to vary.

All of our Fano fits are processed and filtered in the same way as for waveguide bias data, and we repeat the weighted least-squares optimization.

\subsection{Dual Bias}
Lastly, we can predict the tuning behavior under arbitrary bias by feeding the combined results of the waveguide and racetrack bias fits (i.e., the parameters $p_j$) into our complete model in eqs.~\ref{eq:zeta_fit_model}-\ref{eq:delta_fit_model}. We compute the values for bandwidth and detuning under various voltages $V_w$ and $V_r$, yielding the simulated results in main text Fig.~4 and supplemental Fig.~\ref{fig:FigS1_b}. We also then predict the bandwidth and detuning for the specific voltages used in measurement, obtaining the predictions plotted against the simulated data.

\subsection{Error Bars}\label{supp:sec_error}
\begin{figure}[h]
    \centering
    \includegraphics{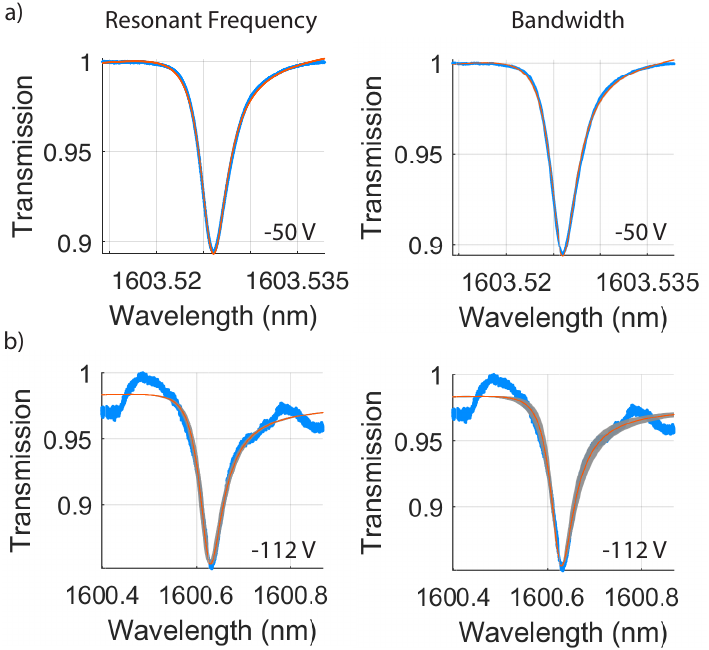}
    \caption{Examples of error-bounded fits for a $50\%$ error bound. We choose such a large error bound to make the resulting parameter adjustments clearly visible to the reader. Blue is data, red is the fit. Grey shading corresponds to all other possible fits as the appropriate fit parameter is adjusted within the $50\%$ error bounds. (a) device 1 resonant frequency adjustment (left) and bandwidth adjustment (right). In this case, the original fit is very accurate, and any changes to the resonant frequency or bandwidth are so minor that we cannot see them on the plots. (b) device 2 resonant frequency adjustment (left) and bandwidth adjustment (right). Here we observe grey shading corresponding to possible fits within the $50\%$ error bounds.}
    \label{fig:supplement_err_bounds}
\end{figure}

For all fits, we determine error bounds, which are used as the weights, $\epsilon$, in the cost function given by eq.~\ref{eq:S_costFunction}. This is most relevant in the high-bandwidth regime in which the mode spectrum begins to merge with the background. In these cases, although a least-squares Fano fit of the mode might have very low fit error, the fit itself may be inaccurate. It is unclear whether the resulting Fano fit yields the true bandwidth and resonant frequency of the mode. Reflecting this, we vary the parameters of the Fano fit independently and calculate the change in the least-squares error for each parameter change. We interpolate the upper and lower values of the parameter for which the Fano fit least-squares error changes by $\pm10\%$. These values then constitute the upper and lower error bounds we plot for the bandwidth. For the detuning, the errorbars are calculated similarly. We determine the error bars on the resonant frequency of the Fano fit and add this error in quadrature with the error in the zero-voltage resonant frequency. Fig.~\ref{fig:supplement_err_bounds} shows examples of fits and errors for waveguide-only bias data in various coupling regimes. Here we plot for an error bound of $\pm50\%$ for visualization purposes, to make the error bounds clearer to the reader. 

The error weights, $\epsilon_i$ in eq.~\ref{eq:S_costFunction}, are given by the average of the upper and lower error bounds on each data point int he tuning curves for bandwidth and detuning.

\section{Asymmetry Analysis}
\begin{figure}[h]
    \centering
    \includegraphics{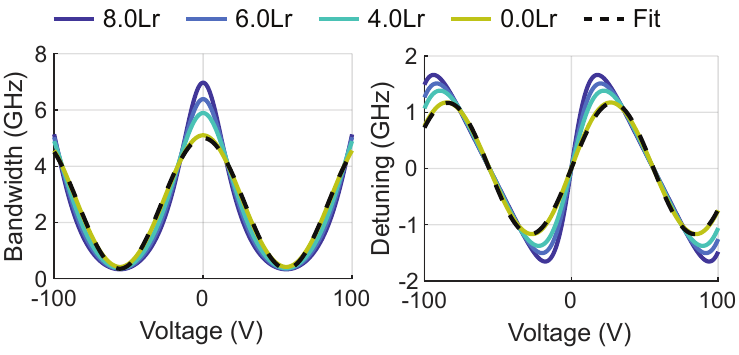}
    \caption{Simulated bandwidth (left) and detuning (right) versus voltage for different lengths of the interferometric waveguide. $L_r$ refers to the half-length of the racetrack resonator. The dashed black lines are fits of cosine functions to the $0L_r$ curves. As the interferometric waveguide vanishes, the tuning asymmetry vanishes and the behavior becomes sinusoidal. The results of the device 1 calibration are used to simulate device performance. Only waveguide bias is ``applied'' to the device in the simulation, and we neglect the quadratic voltage dependence in the phase for this simulation.}
    \label{fig:supplementS6a}
\end{figure}

As discussed in the main text, there are a few physical asymmetries in the device architecture. The first of these is the length difference between the feedback waveguide and the bottom half of the racetrack connecting the two coupling points. We attribute the asymmetric periodicity in the tuning behavior of the device with this length imbalance. In order to explore this, we conduct a number of simulations of the expected tuning behavior for waveguide-only bias, using the fit parameters presented in table \ref{tab:fit_parameters}. The results are shown in Fig.~\ref{fig:supplementS6a}. For a racetrack half-length $L_r$, we sweep the feedback waveguide length. We notice that the asymmetry in the tuning period decreases as the lengths are more closely matched. When the length of the feedback waveguide is negligible ($L_w = 0$), we recover a sinuosoid, as indicated by fitting the simulation data (given by the dashed black line in Fig.~\ref{fig:supplementS6a}).

\begin{figure}[h]
    \centering
    \includegraphics{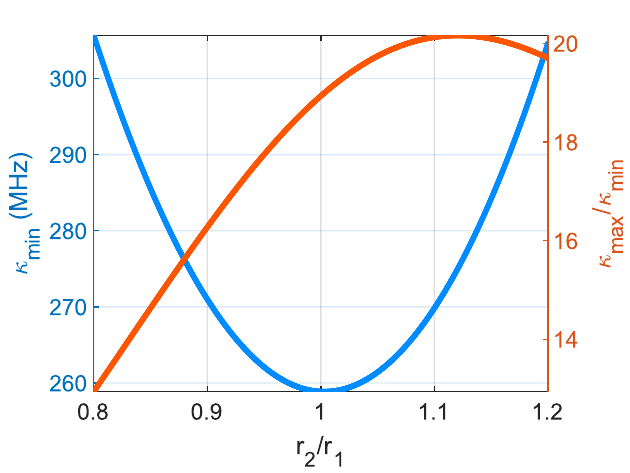}
    \caption{Predicted minimum bandwidth (blue) and bandwidth extinction (BE) ratio (red) as a function of coupler asymmetry. $r_i$ is the amplitude cross-coupling at coupling point $i$, and $r_1$ is kept fixed while $r_2$ is swept to generate the $\kappa_{\text{min}}$ curve. In order to generate the red BE curve, we fix $t_1$, the transmission at coupling point $1$, and sweep $t_2$. Asymmetry in the cross-coupling (or transmission) coefficients is reflected as an increase of intrinsic loss (change in extrinsic loss) in the resonator. We observe an optimal asymmetry point where the bandwidth extinction ratio is maximized.}
    \label{fig:supplementS6b}
\end{figure}

We also explore asymmetry in the coupling points. As depicted in Fig.~\ref{fig:supplementS6b}, when the two coupling points have an unequal amplitude coupling strength, the minimum measureable bandwidth \textit{increases}. Therefore, asymmetry in the coupling points emerges as an additional effective loss in the racetrack resonator. We can reason this by way of imperfect Mach-Zehnder Interferometers (MZIs). In this case, asymmetry in the couplers leads to a decrease in the MZI extinction ratio. Similarly, asymmetry in the couplers in our device leads to incomplete interference of the optical modes in the racetrack and waveguide. Therefore, there will always be some amount of power coupling between the racetrack and the waveguide, which broadens the bandwidth of the racetrack resonance. Furthermore, by keeping one coupler transmission amplitude fixed and sweeping the transmission at the other coupler, we observe a change in the bandwidth extinction ratio. For $t_2/t_1 > 1$, we are simply increasing total transmission and therefore loss from the ring, leading to a growing $\kappa_{\text{max}}$. We observe an optimal point in the asymmetry at which this extinction ratio is maximized given the increasing $\kappa_{\text{max}}$ and increasing $\kappa_{\text{min}}$.

\clearpage

\bibliography{TK02_Paper_FinalBib}
\end{document}